\documentclass[conference]{IEEEtran}
\IEEEoverridecommandlockouts
\usepackage{cite}
\usepackage{booktabs}
\usepackage{multirow}
\usepackage{siunitx}
\usepackage{subcaption}
\usepackage{multicol}
\usepackage{url}
\usepackage{xurl}
\usepackage[pdftex]{graphicx}
\usepackage[export]{adjustbox}
\usepackage{algorithm,algorithmic}
\usepackage{lettrine}
\usepackage{amsmath}
\usepackage{graphicx}
\usepackage[T1]{fontenc}
\usepackage{colortbl}
\usepackage{color}
\usepackage{rotating}
\usepackage{float}
\usepackage{caption}
\usepackage{hhline}
\definecolor{maroon}{cmyk}{0,0.87,0.68,0.32}
\usepackage{cases}
\usepackage{amsfonts}

\usepackage{blindtext}
\usepackage{bm}
\usepackage[switch]{lineno}
\newcommand\amend[1]{\textcolor{blue}}

\usepackage{mathabx}
\usepackage{pifont}
\newlength\myindent
\usepackage{tabu}
\usepackage{authblk}
\usepackage{acronym}

\acrodef{DL}[DL]{Deep Learning}
\acrodef{DNN}[DNN]{Deep Neural Network}
\acrodef{IoMT}[IoMT]{Internet of Medical Things}
\acrodef{FPGA}[FPGA]{Field Programmable Gate Array}
\acrodef{CNN}[CNN]{Convolutional Neural Network}
\acrodef{MDLS}[MDLS]{Memristive Deep Learning System}
\acrodef{EEG}[EEG]{Electroencephalogram}
\acrodef{IT}[IT]{Information Technology}
\acrodef{CAS}[CAS]{Circuits and Systems}
\acrodef{ANN}[ANN]{Artificial Neural Network}
\acrodef{ML}[ML]{Machine Learning}
\acrodef{CHB}[CHB]{Children’s Hospital Boston}
\acrodef{SOP}[SOP]{Seizure Occurrence Period}
\acrodef{SPH}[SPH]{Seizure Prediction Horizon}
\acrodef{FFT}[FFT]{Fast Fourier Transform}
\acrodef{STFT}[STFT]{Short Time Fourier Transform}
\acrodef{WT}[WT]{Wavelet Transform}
\acrodef{PSD}[PSD]{Power Spectral Density}
\acrodef{VMM}[VMM]{Vector Matrix Multiplication}
\acrodef{FPR}[FPR]{False Prediction Rate}
\acrodef{CMOS}[CMOS]{Complementary Metal–Oxide–Semiconductor}
\acrodef{DBS}[DBS]{Deep Brain Stimulation}
\acrodef{VLSI}[VLSI]{Very-large-scale Integration}
\acrodef{LLS}[LLS]{Linear Least Squares}
\acrodef{SVM}[SVM]{Support Vector Machine}
\acrodef{ICA}[ICA]{Independent Component Analysis}
\acrodef{SOPD}[SOPD]{Second-order Difference Plot}
\acrodef{kNN}[kNN]{$k$-nearest neighbors}
\acrodef{ADC}[ADC]{Analog-to-digital Converter}
\acrodef{AUROC}[AUROC]{Area Under the Receiver Operating Characteristic Curve}
\acrodef{1T1R}[1T1R]{1-Transistor 1-Memristor}
\acrodef{NLL}[NLL]{Negative Log Likelihood Loss} 
\acrodef{SOTA}[SOTA]{State-of-the-art}
\acrodef{FH}[FH]{Freiburg Hospital}
\acrodef{ASSPC}[ASSPC]{American Society Seizure Prediction Challenge}
\acrodef{SoC}[SoC]{System on a Chip}
\acrodef{MIT}[MIT]{Massachusetts Institute of Technology}
\acrodef{ADC}[ADC]{Analog to Digital Converter}
\acrodef{RRAM}[RRAM]{Resistive Random Access Memory}
\acrodef{PCM}[PCM]{Phase Change Memory}
\acrodef{STT-MTJ}[STT-MTJ]{Spin-transfer Torque Magnetic Tunnel Junction}
\acrodef{MO}[MO]{Metal Oxide}
\acrodef{FIFO}[FIFO]{First-In First-Out}
\acrodef{mGPU}[mGPU]{Mobile Graphics Processor Unit}
\acrodef{GST}[GST]{Ge$_2$Sb$_2$Te$_5$}
\acrodef{TDM}[TDM]{Time-Division Multiplexing}

\begin{document}
\title{Towards Memristive Deep Learning Systems for Real-time Mobile Epileptic Seizure Prediction}
\author[1]{Corey Lammie\thanks{\hspace{-1em}\rule{3cm}{0.5pt} \newline \textcopyright  \hspace{1pt} 2021 IEEE. Personal use of this material is permitted. Permission from IEEE must be obtained for all other uses, in any current or future media, including reprinting/republishing this material for advertising or promotional purposes, creating new collective works, for resale or redistribution to servers or lists, or reuse of any copyrighted component of this work in other works.}}
\author[2]{Wei Xiang}
\author[1]{Mostafa Rahimi Azghadi}
\affil[1]{College of Science and Engineering, James Cook University, Queensland 4814, Australia \authorcr Email:\{corey.lammie, mostafa.rahimiazghadi\}@jcu.edu.au}
\affil[2]{Department of Computer Science and Information Technology, La Trobe University, Victoria 3086, Australia \authorcr Email: w.xiang@latrobe.edu.au}

\maketitle

\begin{abstract}
The unpredictability of seizures continues to distress many people with drug-resistant epilepsy. On account of recent technological advances, considerable efforts have been made using different hardware technologies to realize smart devices for the real-time detection and prediction of seizures. In this paper, we investigate the feasibility of using \acp{MDLS} to perform real-time epileptic seizure prediction on the edge. Using the \textit{MemTorch} simulation framework and the \ac{CHB}-\ac{MIT} dataset we determine the performance of various simulated \ac{MDLS} configurations. An average sensitivity of 77.4\% and a \ac{AUROC} of 0.85 are reported for the optimal configuration that can process \ac{EEG} spectrograms with 7,680 samples in 1.408ms while consuming 0.0133W and occupying an area of 0.1269mm$^2$ in a 65nm \ac{CMOS} process.
\end{abstract}

\color{black}

\section{Introduction}
\lettrine{T}{he} backbone of smart healthcare is the \ac{IoMT}, which is an amalgamation of medical devices and applications that connect through the internet to healthcare \ac{IT}~\cite{Dimitrov2016} to overcome the shortcomings of traditional healthcare. The \ac{IoMT} has the potential to give rise to many medical applications, including mobile epileptic seizure prediction, which is the primary focus of this paper. 

\ac{IoMT} edge devices can be used to perform computations locally, reducing latency and alleviating privacy concerns when sensitive medical data is processed. Moreover, they can be used to realize closed-loop systems, which are highly desirable for patient monitoring and treatments~\cite{azghadi2020hardware}. In Fig.~\ref{fig:block_diagram}, we depict three different application scenarios of our proposed seizure prediction system. To enable such a smart \ac{DL}-based system to operate in real-time at the power-constrained edge, \ac{RRAM}-based in-memory \ac{DL} computing architectures~\cite{rahimi2020complementary} could be used~\cite{azghadi2020hardware}. In this paper, we investigate the feasibility of using \acp{MDLS} to perform real-time epileptic seizure prediction at the edge to enable a mobile solution. Our specific contributions are as follows:

\begin{figure}[!t]
\centering
\includegraphics[width=0.425\textwidth]{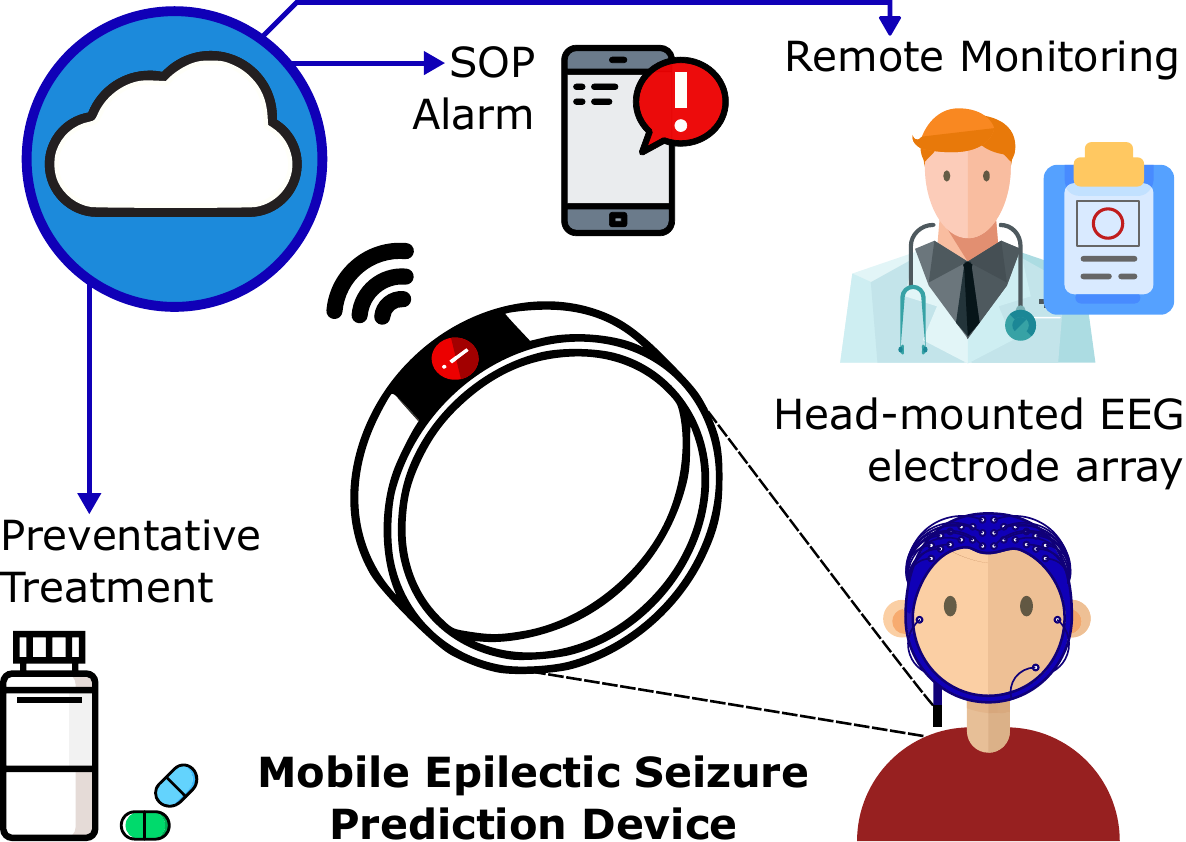}
\caption{
Application scenarios of the proposed system, which is able to facilitate a variety of treatment types.
}
\label{fig:block_diagram}
\end{figure}

\begin{figure*}[!t]
\centering
\includegraphics[width=1\textwidth]{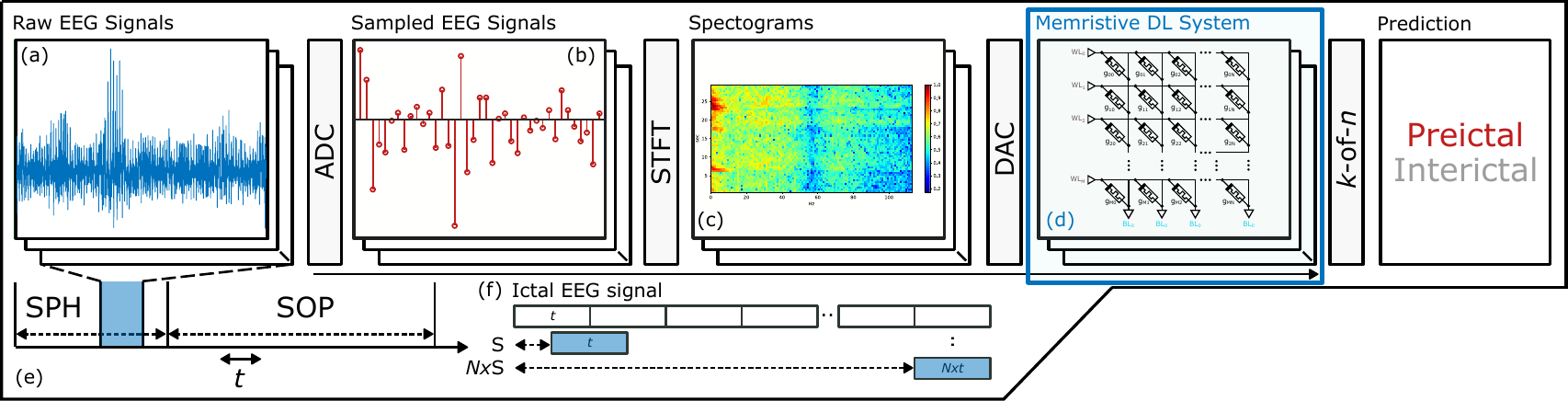}
\caption{
A simplified block diagram (a--e) of the proposed system and (f) a depiction of the methodology used to generate synthetic preictal samples. Raw \ac{EEG} signals (a) are measured using several electrodes, which are (b) sampled using \acp{ADC}. \ac{CMOS} circuits~\cite{Tsai2012,Lin2020} are used to filter and generate (c) spectrograms for each window, $t$, using the discrete \ac{STFT}. A (d) \ac{MDLS} is used to perform in-memory computation to predict the state of future samples that are to occur in the \ac{SOP} (preictal or interictal) during the \ac{SPH}. During training, (f) synthetic preictal samples are generated to balance the number of preictal and interictal samples. Extra preictal samples are generated by sliding a 30 second window along the time axis at every step, $S$, over preictal signals~\cite{Truong2018}.
}
\label{fig:approach}
\end{figure*}

\begin{enumerate}
\item We are the first to investigate an in-memory \ac{DL} approach to epileptic seizure prediction;
\item We explore a variety of weight-representation schemes while accounting for some device nonidealities, and compare the performance of our approach to other \ac{DL} approaches;
\item We determine the power and area requirements for the optimal configuration, and investigate its feasibility for eventual hardware realization. 
\end{enumerate}

\section{Related Work}
To the best of our knowledge, all existing hardware implementations tasked for epileptic seizure detection and prediction have been realized using \ac{FPGA}, \ac{CMOS} and \ac{VLSI} technologies. Most existing hardware implementations detect epileptic seizures using traditional \ac{ML} algorithms such as \ac{LLS}~\cite{6107718}, \acp{SVM}~\cite{8467308}, and \ac{kNN}~\cite{6996043}. We refer the reader to~\cite{Alotaiby2014} for a comprehensive survey of epileptic seizure detection and prediction systems. While \acp{ANN} have previously been used for epileptic seizure detection~\cite{5639541} and prediction~\cite{Daoud2020} on \ac{FPGA}, no previous work has investigated the use of memristors for the detection or prediction of epileptic seizures using \ac{DL}, which could drastically improve the performance on the \ac{IoMT} edge.

\section{Preliminaries}
\subsection{Seizure Forecasting Systems}
There is emerging evidence~\cite{Smith2005} that the temporal dynamics of brain activity of people with epilepsy can be classified into 4 states: interictal (between seizures, or baseline), preictal (prior to seizure), ictal (seizure), and post-ictal (after seizures). Seizure forecasting or predictive systems aim to classify the preictal brain state.

\subsection{Memristive DL Systems}
Memristive devices can be arranged within crossbar architectures to perform \acp{VMM} in-memory, in $\mathcal{O}(1)$~\cite{Hu2018}, which are used extensively in forward and backward propagations within \acp{CNN} to compute the output of fully connected and unrolled convolutional layers. Scaled weight matrices can either be represented using two crossbars per layer, $g_{\text{pos}}$ and $g_{\text{neg}}$, to represent positive and negative weights, respectively, or using a singular crossbar per layer with current mirrors, so that the effective conductance of each device is offset by a fixed value, $g_m$, that can be determined using (\ref{eq:g_m})~\cite{Lammie2020MemTorchAO}   
\begin{equation}\label{eq:g_m}
g_m = -2 / (\bar{R_{\textnormal{ON}}} + \bar{R_{\textnormal{OFF}}}),
\end{equation}

\noindent where crossbar column currents can be multiplied by a layer-specific scaling parameter, $K$, to determine layer outputs. When a single device is used to represent each parameter, constant currents to mirror can easily be realized using a diode-connected NMOSFET by adjusting the NMOSFET channel width so that it has a passive conductance $g_m$. Given scalability issues, large crossbars can be split into smaller ones, referred to as either modular crossbar arrays, or crossbar tiles~\cite{Mountain2018} to compute the output of linear and convolutional layers with a large number of weights. 

\section{Proposed System}
A simplified block diagram of the proposed system is provided in Fig.~\ref{fig:approach}. We confine the scope of this paper solely to the memristive DL system component depicted in Fig.~\ref{fig:approach}(d), and only consider instances where learning is performed offline.

\subsection{Network Architecture}
The network architecture used is summarized in Table \ref{network_architecture}, where $n$ is the number of electrodes that are used to sample \ac{EEG} signals, $t$ is the window size in seconds, and $p$ can be determined using (\ref{eq:p})
\begin{equation}\label{eq:p}
p = tf_s/k_s = 2t,
\end{equation}

\noindent where $k_s$ denotes the number of overlapped samples, which for all cases in this paper is fixed to 128, i.e., half the sampling frequency, $f_s$. Batch normalization and the ReLU activation function is applied to the output of all convolutional layers and the first fully connected layer. The output of the last fully connected layer is fed through a Softmax activation function. In contrast to other architectures used in related works~\cite{Truong2018,Kiral-Kornek2018,8239676}, our architecture uses only linear, 2d-convolutional, max pooling, and batch normalization layers. 

\begin{table}[!b]
\centering
\caption{Network architecture employed. For each convolutional and pooling layer, $f$ is the number of filters, $k$ determines the filter size, and $s$ denotes the stride length. For each fully connected layer $N$ denotes the number of output neurons.}
\begin{tabu} to 0.5\textwidth {p{0.3\textwidth}X[r]}
\toprule
\multicolumn{2}{c}{\textbf{Input ($n \times p \times 114$)}} \\
\midrule
\textbf{Layer} & \textbf{Output Shape} \\
\midrule
Convolutional, $f = 16, k=(5, 5), s=(2, 2)$  & $(16 \times [p-3]/2 \times 55)$  \\
Max Pooling, $k=(2, 2)$  & $(16 \times [p-3]/4 \times 27)$  \\
Convolutional, $f = 32, k=(3, 3), s=(1, 1)$  & $(32 \times [p-11]/4 \times 25)$  \\
Max Pooling, $k=(2, 2)$  & $(32 \times [p-11]/8 \times 12)$  \\
Convolutional, $f = 64, k=(3, 3), s=(1, 1)$  & $(64 \times [p-27]/8 \times 10)$  \\
Max Pooling, $k=(2, 2)$  & $(64 \times [p-27]/16 \times 5)$  \\
Fully Connected, $N = 256$  & $(256)$  \\
Fully Connected, $N = 2$  & $(2)$  \\
\bottomrule
\end{tabu}\label{network_architecture}
\end{table}

\begin{table*}[!t]
\centering
\captionsetup{justification=centering}
\caption{Patient information and performance metrics across all folds for our trained conventional \acp{CNN} and their equivalent \acp{MDLS} adopting a double-column parameter-representation scheme.}
\begin{tabu} to \textwidth {XXp{0.15\textwidth}XX[c]X[c]X[c]X[c]}
\toprule
Patient & Seizures & Interictal Duration (h) & S & Accuracy (\%) & Sensitivity (\%) & AUROC & FPR (/h) \\ 
\midrule
1 & 7 & 17.0 & 7.122 & 94.36$\pm$0.99  & 79.72$\pm$0.01  & 0.97$\pm$0.01  & 01.4$\pm$0.2  \\
2 & 3 & 22.9 & 1.684 & 94.36$\pm$1.40  & 94.31$\pm$0.01  & 0.97$\pm$0.01  & 01.6$\pm$0.4  \\
5 & 5 & 13.0 & 5.060 & 74.16$\pm$1.82  & 80.42$\pm$0.01  & 0.85$\pm$0.01  & 08.0$\pm$0.5  \\
19 & 3 & 24.9 & 1.687 & 96.33$\pm$0.66  & 54.72$\pm$0.00  & 0.50$\pm$0.00  & 20.6$\pm$0.0  \\
23 & 5 & 3.0 & 7.244 & 94.43$\pm$3.08  & 79.51$\pm$0.02  & 0.96$\pm$0.02  & 07.8$\pm$4.4  \\
\bottomrule
\end{tabu}\label{table:performance}
\end{table*}

\begin{figure*}[!t]
\centering
\includegraphics[width=1\textwidth]{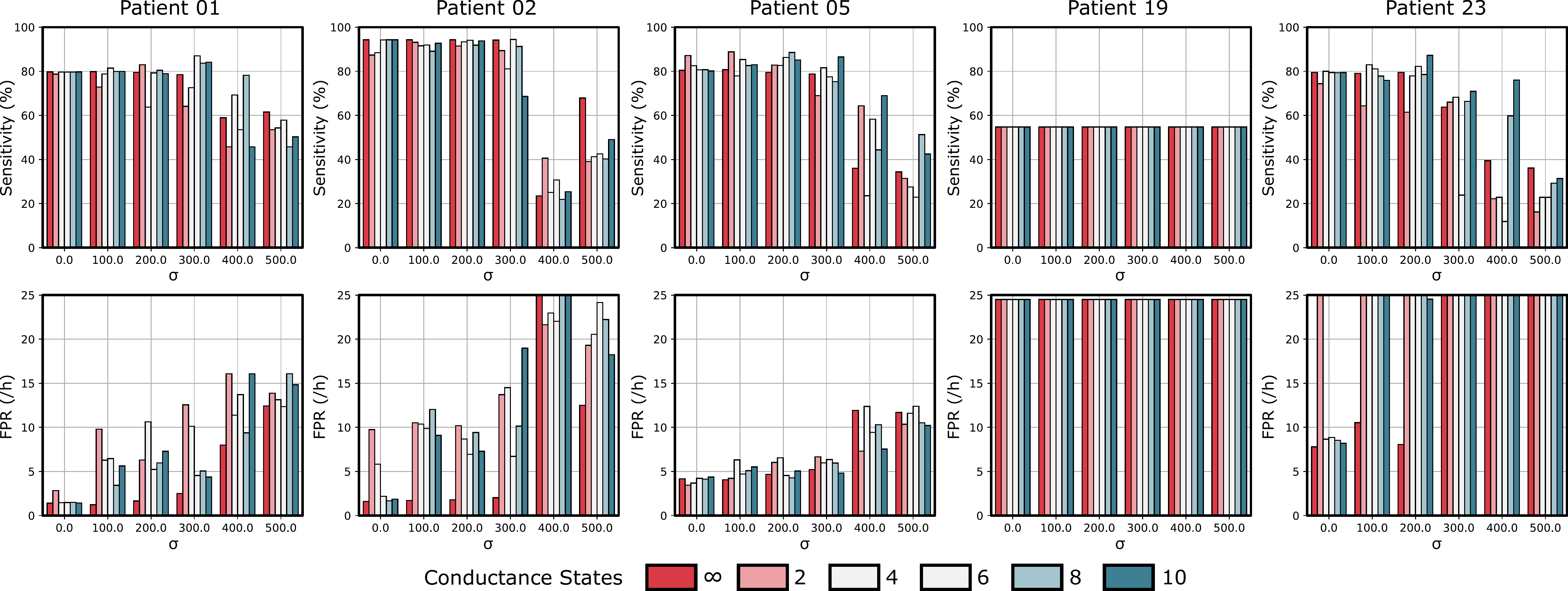}
\caption{The average sensitivity and \ac{FPR} across all 5 folds for simulated double column \ac{MDLS} configurations.}
\label{fig:results}
\end{figure*}

\subsection{Training and Validation Datasets}
For training and validation of our \ac{MDLS}, we used the \ac{CHB}-\ac{MIT}~\cite{shoeb2010application} dataset, which consists of \ac{EEG} recordings from 22 pediatric subjects with intractable seizures. For our preliminary study reported in this paper, 5 random patients were chosen. We leave evaluation using all subjects from the \ac{CHB}-\ac{MIT} and other datasets, such as the \ac{ASSPC}, to more exhaustive future works. 

\subsection{Preprocessing Steps}
Within the \ac{CHB}-\ac{MIT} dataset, there are instances where multiple seizures occur in close proximity to each other. For seizure prediction, we are interested in predicting leading seizures. Consequently, seizures that occur $\leq T$ minutes after a previous seizure are not considered, where $T$ denotes the \ac{SOP}. All time-series \ac{EEG} signals are translated into time-frequency signals using \acp{STFT} with a window length of $t$ seconds (Fig.~\ref{fig:approach}(e-f)). Similarly to~\cite{Truong2018}, power line noise was removed by excluding components in the frequency ranges of 57–63 Hz and 117–123 Hz. The DC component (at 0 Hz) and components of frequencies above 114 Hz were also removed. 

\subsection{Training and Validation Methodologies}
On account of the large class imbalance between preictal and interictal samples, we use an overlapped sampling technique, which was originally proposed in~\cite{Truong2018}, to train the adopted network architecture. This is depicted in Fig.~\ref{fig:approach}(f). Extra preictal samples are generated by sliding a $t$ second window along the time axis at every step, $S$, over preictal samples, which is chosen so that there are a similar number of samples per class (preictal or interictal). The \ac{NLL} function was used in conjunction with the DiffGrad optimization algorithm, which has been shown to outperform other optimizers~\cite{8939562}, to train the networks with an initial learning rate of $1e^{-4}$ and batch size of $256$ for 50 epochs, when performance stagnated. For a correct prediction, a seizure onset must be after the \ac{SPH} and within the \ac{SOP}, as depicted in Fig.~\ref{fig:approach}. The metrics used to test the proposed approach are the accuracy, sensitivity, \ac{AUROC}, and the \ac{FPR}, as shown in Table \ref{table:performance}. For each subject, performance is reported using $k=5$ stratified K-fold cross validation, where synthetic samples are discarded during evaluation. All implementations adopted the following parameters: $T=30$ minutes, $t=30$ seconds, and a \ac{SPH} of 35 minutes.

\section{Performance Evaluation}
The MemTorch~\cite{Lammie2020MemTorchAO} simulation framework was used to simulate \ac{RRAM} devices during inference using the VTEAM~\cite{7110565} model. Performance metrics for our trained conventional and equivalent \ac{MDLS} are reported in Table \ref{table:performance}. When predicting \ac{EEG} seizures, it is common to have isolated false positives during interictal periods~\cite{Truong2018}. In recent works, discrete-time Kalman filters and least-$k$-prediction post-processing techniques have been adopted, however, they introduce a significant hardware overhead. 

In Fig.~\ref{fig:results}, we report the sensitivity and \ac{FPR} for all simulated configurations adopting a double column weight-representation scheme, as the performance of all configurations adopting a single column weight-representation scheme is insignificant. Consequently, we determine the optimal configuration to be a network adopting a double column weight-representation scheme. We attribute the high \ac{FPR} for all trained networks and simulated configurations to the omission of any data post-processing, which is out of the scope of this paper. For all devices, $\bar{R_{\text{ON}}} = 100\Omega$ and $\bar{R}_{\text{OFF}} = 2,500\Omega$~\cite{Yalon2012}. Two non-ideal device characteristics were modelled: device-to-device variability, and a finite number of discrete conductance states. Device-device variability was introduced stochastically by sampling $R_{\text{ON}}$ and $R_{\text{OFF}}$ for each device from a normal distribution with $\bar{R_{\text{ON}}} = 100\Omega$ and $\sigma$, and $\bar{R_{\text{OFF}}} = 2,500$ and $2\sigma$, as $\bar{R_{\text{OFF}}} \gg \bar{R_{\text{ON}}}$~\cite{Lammie2020MemTorchAO}, for $\sigma$ = 0--500. As it has been demonstrated that the spacing between states is not critical~\cite{Mehonic2019}, we simulated devices with between 2–-10 uniformly distributed conductance states. 

From Fig.~\ref{fig:results}, it can be observed that for patients 1, 2, 5, and 23, the sensitivity and \ac{FPR} decreased when the number of conductance states decreased and device-device variability increased. Interestingly, while the number of finite conductance states did not have a large influence on the reported sensitivity and \ac{FPR} for these patients, device-device variability did. The sensitivity has a relatively sudden transition period at $\sigma > 300$, when the distributions of $R_\text{\text{ON}}$ and $R_\text{\text{OFF}}$ overlapped, causing the sensitivity to abruptly decrease. Conversely, the \ac{FPR} was much more sensitive to device-device variability. It is noted that, for patient 19, we report an average accuracy of 96.33\% and sensitivity of 54.72\%. While this result cannot be clearly explained, it is not uncommon in literature, and other \ac{DL} works~\cite{Truong2018,8239676} using the same dataset also report a high accuracy and low sensitivity, near $50\%$ for the same patient. 

\begin{table}[!b]
\centering
\caption{Power, area, and latency requirements of the optimal configuration using 128$\times$128 crossbar tiles for TDM and parallelized implementations (Imp.).}
\begin{tabu} to 0.4\textwidth {XXXXX}
\toprule
\textbf{Imp.} & \multicolumn{1}{r}{\textbf{Power (W)}} & \multicolumn{1}{r}{\textbf{Area (mm$^2$)}} & \multicolumn{1}{r}{\textbf{Latency (ms)}} & \multicolumn{1}{r}{\textbf{Energy (mJ)}} \\
\midrule
TDM & 0.0133 & 0.1269 & 1.408 & 0.0187\\
Parallelized & 1.7 & 8.5089 & 0.011 & 0.0187\\
\bottomrule
\end{tabu}\label{table:power_area}
\end{table}

\subsection{Comparison to Other \ac{DL} Models}
Since previous related works~\cite{Truong2018,8239676} do not use a consistent testing methodology, we can only roughly compare our results to them using the sensitivity and \ac{FPR} metrics from patients 1, 2, 5, 19, and 23 from Table \ref{table:performance}. Ref.~\cite{Truong2018} and~\cite{8239676} report total sensitivities of $81.2\%$ and $87.8\%$, and \acp{FPR} of 0.16/hr and 0.14/hr, respectively. In~\cite{8239676} clinical considerations were discarded and a zero \ac{SPH} was used. Consequently, the reported performance is likely inflated. Nevertheless, as we did not perform any data post-processing, compared to both works, all of our networks have significantly larger \acp{FPR}. In Table \ref{table:performance}, we report an average sensitivity of $77.74\%$, which is lower than that reported in~\cite{Truong2018} and~\cite{8239676}. Our result is still significant, because we use 2d-convolutional layers, max pooling, and fully-connected layers, and perform minimal data processing, while~\cite{Truong2018} used 3d-convolutional layers and~\cite{8239676} performed hyper-parameter optimization to obtain the lowest average validation loss over a 10 fold cross-validation. 

\subsection{Power, Area, and Delay Analysis}
To determine the power and area requirements as well as the latency, which dictates the inference time of the optimal configuration, we map each layer of our deep network to modular 128$\times$128 crossbar tiles with no shared weights between layers using parameters for 65nm technology from~\cite{Wang2019}. The area and power of each \ac{ADC} (8-bit) is, therefore, calculated to be 3$\times$10$^{-3}$mm$^2$ and 2$\times$10$^{-4}$W, and the area of each RRAM cell is estimated to be 1.69$\times$10$^{-7}$mm$^2$. During inference, we assume constant operation at $V=0.3$V per active cell, the largest voltage used to encode inputs, and an average cell resistance of $(\bar{R}_{\text{OFF}} + \bar{R}_{\text{ON}}) / 2$. All \acp{ADC} are assumed to operate at 5 MHz, and the number of tiles used for each network is assumed to be the exact number required to balance the latency among layers. \ac{RRAM} read latency is considered negligible compared to \ac{ADC} readout.

Table~\ref{table:power_area} shows the power, area, latency, and energy of our optimal configuration for configurations where samples are continuously fed to the network from a \ac{FIFO} buffer. We compare requirements for implementations for which each tile contains one \ac{ADC}, and \ac{TDM} is used to read out column currents (denoted \textit{\ac{TDM}}), and for which each tile contains one \ac{ADC} per column to read out column currents in parallel (denoted \textit{Parallelized}). Given the large window length used, further duplication of crossbar tiles to improve throughput was deemed unnecessary. 

\section{Conclusion}
We investigated the potential of memristors to contribute to the design of a \ac{DL}-based seizure prediction device. 
Our findings demonstrate that \ac{MDLS} holds great promise for developing a compact epileptic seizure prediction architecture capable of low-power and real-time mobile operation. Our optimal configuration exhibits comparable performance to existing \ac{DL} works in the literature while consuming significantly less power than current \acp{mGPU} and edge processors~\cite{azghadi2020hardware}. In future, the longevity and reliability of such a system should be properly investigated.

\bibliographystyle{IEEEtran}
\bibliography{References}
\end{document}